\documentclass[prd,aps,showpacs,epsf,floats]{revtex4}
\usepackage{amsfonts}
\usepackage{amsmath}

\input{tcilatex}
\begin{document}

\title{Can Magnetic Monopoles And Massive Photons Coexist In The Framework
Of The Same Classical Theory?}
\author{C. Cafaro${}$}
\email{carlocafaro2000@yahoo.it}
\affiliation{Department of Physics, State University of New York at Albany-SUNY,1400
Washington Avenue, Albany, NY 12222, USA}
\author{S. Capozziello}
\email{capozziello@na.infn.it (Corresponding Author)}
\affiliation{Dipartimento di Scienze Fisiche, Universit\`{a} di Napoli "Federico II", Via
Cinthia, 80126 Napoli, Italy}
\author{Ch. Corda}
\email{christian.corda@ego-gw.it}
\affiliation{INFN Sezione di Pisa and University of Pisa, Via F. Buonarroti 2, 56127
Pisa, Italy; European Gravitational Observatory (EGO), Via E. Amaldi, 56021
Cascina (PI), Italy}
\author{S. A. Ali}
\email{alis@alum.rpi.edu}
\affiliation{Department of Physics, State University of New York at Albany-SUNY,1400
Washington Avenue, Albany, NY 12222, USA}

\begin{abstract}
It is well known that one cannot construct a self-consistent quantum field
theory describing the non-relativistic electromagnetic interaction mediated
by massive photons between a point-like electric charge and a magnetic
monopole. We show that, indeed, this inconsistency arises in the classical
theory itself. No semi-classic approximation or limiting procedure for $%
\hbar \rightarrow 0$ is used. As a result, the string attached to the
monopole emerges as visible also if finite-range electromagnetic
interactions are considered in classical framework.
\end{abstract}

\maketitle


In his classical works, Dirac showed that the existence of a magnetic
monopole would explain the electric charge quantization \cite{dirac}. This
is known as the Dirac quantization rule. There exist various arguments based
on quantum mechanics, theory of representations, topology and differential
geometry on behalf of Dirac's rule \cite{jackiw, Nash}. Dirac's formulation
of magnetic monopoles takes into account a singular vector potential. Other
approaches exist where two non-singular vector potentials, related through a
gauge transformation, are used \cite{WuYang1, WuYang2}. Finite-range
electrodynamics is a theory with non-zero photon mass. It is an extension of
the standard theory and is fully compatible with experiments. The existence
of Dirac's monopole in massless electrodynamics is compatible with the above
quantization condition if the string attached to the monopole is invisible.
The quantization condition can be obtained either with the help of gauge
invariance or angular momentum quantization. In massive electrodynamics,
both these approaches are no longer applicable \cite{ignatiev}. These
conclusions are formulated in a quantum framework which is a quantized
version of the classical one. The Hamiltonian formulation and the problems
involved in quantization of Dirac's theory of monopoles have been
extensively discussed in the past and is still an active field of research 
\cite{Balachandran, Nesterov}. Major work on the quantum field theory of
magnetic charges has been developed by Schwinger \cite{Schwinger,
Schwinger2, Schwinger3} and Zwanziger \cite{Zwanziger}. Recent work on
constructing a satisfactory classical relativistic framework for massive
electrodynamics and magnetic monopoles from a geometrical point of view has
been considered in \cite{israelit1, israelit2}. A complete update on the
experimental and theoretical status of monopoles is presented in \cite%
{Milton}.

In this letter we consider the problem of constructing the static limit of a
consistent classical, non-relativistic electromagnetic theory describing a
point-like electric particle with charge $e$\ and mass $m$\ moving in the
field of a fixed composite monopole of charge $e_{m}$, where their mutual
interaction is mediated by massive carrier gauge fields. The total magnetic
field $\vec{B}$\ is comprised of point-like magnetic charge, a semi-infinite
string along the negative $z$-axis and diffuse magnetic field contributions.
We impose that the electrically charged particle must never pass through the
string ("Dirac-veto") \cite{Brandtand}\ and therefore the motion of the test
charged particle is constrained to region of motion $R^{+}:=\left\{ \left( r%
\text{, }\theta \text{, }\varphi \right) :r\in 
\mathbb{R}
_{0}^{+}\text{, }\theta \in \lbrack 0\text{, }\pi )\text{, }\varphi \in
\lbrack 0\text{, }2\pi ]\right\} $. It is known that no spherically
symmetric diffuse magnetic field solutions are allowed in Maxwell's
classical electrodynamics with massive photons and magnetic monopoles \cite%
{ignatiev}. Requiring the theory presented here be endowed with a
well-defined canonical Poisson bracket structure, it is shown that the total
angular momentum is the generator of rotations. Furthermore, by demanding
proper transformation rules under spatial rotations for the allowed magnetic
vector field solutions, it is shown that only spherically symmetric diffuse
magnetic fields satisfy the Lie algebra of the system. This leads to
conclude that the permitted solutions to the generalized Maxwell theory are
incompatible with the Lie algebra of the Hamiltonian formulation. As a
consequence, any quantization procedure applied to this classical theory
would lead to an inconsistent quantum counterpart.


Maxwell's equations with non-zero photon mass and magnetic charge follow
from a standard variational calculus \cite{CafaroAli, Kyriakopoulos} of the
Maxwell-Proca-Monopole action functional. The field equations for the
electromagnetic 4-vector potential $A_{\mu }$\ together with the Bianchi
identities and Lorenz gauge condition $\partial _{\mu }A^{\mu }=0$, lead to
the generalized Maxwell equations in three-dimensions:%
\begin{equation}
\vec{\nabla}\cdot \vec{E}=4\pi \rho _{e}-m_{\gamma }^{2}A_{0}\text{, }\vec{%
\nabla}\times \vec{E}=-c^{-1}\partial _{t}\vec{B}-4\pi c^{-1}\vec{j}_{m}%
\text{,}
\end{equation}%
\begin{equation}
\vec{\nabla}\cdot \vec{B}=4\pi \rho _{m}\text{, }\vec{\nabla}\times \vec{B}%
=4\pi c^{-1}\vec{j}_{e}+c^{-1}\partial _{t}\vec{E}-m_{\gamma }^{2}\vec{A}%
\text{,}
\end{equation}%
where $m_{\gamma }=\frac{\omega }{c}$ and $\omega $ is the frequency of the
photon. In absence of electric fields, charges and currents, as well as the
absence of magnetic current, the static monopole-like solution of this
system is,%
\begin{equation}
\vec{B}=\vec{B}^{(Dirac)}+\vec{B}_{\gamma }
\end{equation}%
where $\vec{B}^{(Dirac)}$ is the standard Dirac magnetic field,%
\begin{equation}
\vec{B}^{(Dirac)}=\frac{e_{m}}{r^{2}}\widehat{r}
\end{equation}%
whose divergence and curl are given by,%
\begin{equation}
\vec{\nabla}\cdot \vec{B}^{(Dirac)}=\ 4\pi e_{m}\delta ^{(3)}(\vec{r})\text{
and\ }\vec{\nabla}\times \vec{B}^{(Dirac)}=0\text{.}
\end{equation}%
The diffuse magnetic field $\vec{B}_{\gamma }(\vec{r})$ is given by the
following general expression,%
\begin{equation}
\vec{B}_{\gamma }(\vec{r})=b_{\gamma }^{(1)}(r\text{, }\widehat{n}\cdot \vec{%
r})\vec{r}+b_{\gamma }^{(2)}(r\text{, }\widehat{n}\cdot \vec{r})\widehat{n}
\end{equation}%
where $b_{\gamma }^{(1)}$ and $b_{\gamma }^{(2)}$ are general scalar field
functions and $\widehat{n}$ is a unitary vector along the monopole string.
The magnetic field $\vec{B}_{\gamma }(\vec{r})$ is such that,%
\begin{equation}
\vec{\nabla}\cdot \vec{B}_{\gamma }=0\;\text{and, }\vec{\nabla}\times \vec{B}%
_{\gamma }=-m_{\gamma }^{2}(\ \vec{A}^{(Dirac)}+\vec{A}_{\gamma })\text{.}
\end{equation}%
The vector $\vec{A}^{(Dirac)}$\ is the standard singular vector potential
representing the field of a fixed monopole,%
\begin{equation}
\vec{A}^{(Dirac)}(\vec{r})=\frac{e_{m}}{r^{2}}\frac{\sin (\theta )}{1+\cos
(\theta )}(\widehat{n}\times \vec{r})\text{, }\theta \neq \pi \text{,}
\end{equation}%
with semi-infinite singularity line oriented along the negative $z$-axis,
where $e_{m}$\ is the magnetic charge. The vector potential $\vec{A}_{\gamma
}\left( \vec{r}\right) $\ is given by the following general expression,%
\begin{equation}
\vec{A}_{\gamma }\left( \vec{r}\right) =e_{m}m_{\gamma }^{2}f_{\gamma
}(m_{\gamma }r,m_{\gamma }\vec{r}\cdot \widehat{n})(\widehat{n}\times \vec{r}%
)
\end{equation}%
where $f_{\gamma }$\ is a generic scalar field function. Because of the
second equation in $(7)$, it is clear that no spherically symmetric diffuse
magnetic field solutions are allowed, that is to say, solutions like%
\begin{equation}
\vec{B}_{\gamma }(\vec{r})=B_{\gamma }(r)\widehat{r}
\end{equation}%
are not allowed.


On the other hand, it is known that the classical non-relativistic theory
describing the massless electromagnetic scattering of an electric charge
from a fixed magnetic monopole does have a Hamiltonian formulation \cite%
{goldhaber}. With this result in mind, let us consider the classical
non-relativistic theory describing a point-like electric particle with
charge $e$\ and mass $m$\ moving in the field of a fixed monopole of charge $%
e_{m\text{ }}$, but let us suppose that the electromagnetic interaction is
mediated by massive photons. The total magnetic field $\vec{B}$\ is
comprised of the point-like magnetic charge, string and diffuse magnetic
field contributions%
\begin{equation}
\vec{B}=\vec{B}^{(Dirac)}+\vec{B}_{\gamma }=\left[ \vec{\nabla}\times \vec{A}%
^{\left( Dirac\right) }+e_{m}\vec{f}\left( \vec{r}\right) \right] +\vec{%
\nabla}\times \vec{A}_{\gamma }=\vec{\nabla}\times \vec{A}+e_{m}\vec{f}%
\left( \vec{r}\right) \text{, }\vec{A}=\vec{A}^{\left( Dirac\right) }+\vec{A}%
_{\gamma }
\end{equation}%
where%
\begin{equation}
\left\Vert \vec{f}\left( \vec{r}\right) \right\Vert =4\pi \delta \left(
x\right) \delta \left( y\right) \Theta \left( -z\right) =\frac{4\pi }{r^{2}}%
\frac{\delta \left( \theta \right) \delta \left( \varphi \right) }{\sin
\theta }\Theta \left( -\cos \theta \right)
\end{equation}%
is the string function having support only along the line $\widehat{n}=-%
\widehat{z}$\ and passing through the origin while $\Theta $\ is the
Heaviside step function.

The classical Newtonian equation of motion describing this system is%
\begin{equation}
m\frac{d^{2}\vec{r}}{dt^{2}}-\frac{e}{c}\frac{d\vec{r}}{dt}\times \left( 
\vec{\nabla}\times \vec{A}\right) -\frac{ee_{m}}{c}\frac{d\vec{r}}{dt}\times 
\vec{f}\left( \vec{r}\right) =0\text{.}
\end{equation}%
The Hamiltonian that gives rise to the above equations of motion reads%
\begin{equation}
H_{total}(\vec{p}\text{, }\vec{r})=\frac{(\vec{p}-\frac{e}{c}\vec{A})^{2}}{2m%
}+H_{string}\text{, }H_{string}=-\frac{ee_{m}}{c}\int \left( \frac{d\vec{r}}{%
dt}\times \vec{f}\left( \vec{r}\right) \right) \cdot d\vec{r}\text{.}
\end{equation}%
We impose that the electrically charged particle must never pass through the
string ("Dirac-veto") and therefore the classical equation of motion in the
allowed region of motion $R^{+}:=\left\{ \left( r\text{, }\theta \text{, }%
\varphi \right) :r\in 
\mathbb{R}
_{0}^{+}\text{, }\theta \in \lbrack 0\text{, }\pi )\text{, }\varphi \in
\lbrack 0\text{, }2\pi ]\right\} $\ is given by%
\begin{equation}
m\frac{d^{2}\vec{r}}{dt^{2}}-\frac{e}{c}\frac{d\vec{r}}{dt}\times \left( 
\vec{\nabla}\times \vec{A}\right) =0\text{.}  \label{reduced-EOM}
\end{equation}%
The restricted Hamiltonian associated with (\ref{reduced-EOM}) is given by 
\begin{equation}
H(\vec{p}\text{, }\vec{r})=\frac{(\vec{p}-\frac{e}{c}\vec{A})^{2}}{2m}=\frac{%
\left( \vec{P}\cdot \widehat{r}\right) ^{2}}{2m}+\frac{\left( \vec{P}\cdot 
\widehat{\theta }\right) ^{2}}{2m}=\frac{\left( \vec{P}\cdot \widehat{r}%
\right) ^{2}}{2m}+\frac{\vec{L}^{2}}{2mr^{2}}=\frac{\left( \vec{P}\cdot 
\widehat{r}\right) ^{2}}{2m}+\frac{(\vec{J}^{2}-\vec{s}^{2})}{2mr^{2}}
\end{equation}%
where $\vec{p}=m\frac{d\vec{r}}{dt}+\frac{e}{c}\vec{A}$\ is the canonical
momentum vector, $\vec{P}=\vec{p}-\frac{e}{c}\vec{A}=m\frac{d\vec{r}}{dt}$\
is the kinetic momentum vector, $\vec{L}=\vec{r}\times \vec{P}$\ is the
orbital angular momentum of the system and $\vec{J}=\vec{L}+\vec{s}$\ is the
total angular momentum such that $\vec{J}\cdot \vec{s}=0$\ where 
\begin{equation}
\vec{s}=(4\pi c)^{-1}\int \left[ \vec{r}\times \left( \vec{E}\times \vec{B}%
\right) \right] d^{3}\vec{r}=\vec{s}_{massless}+\frac{e}{4\pi c}\int d\vec{r}%
\vec{r}\times \left[ \frac{\vec{r}}{r^{3}}\times \vec{B}_{\gamma }\left( 
\vec{r}-\vec{R}\right) \right] \text{,}
\end{equation}%
with $\vec{s}_{massless}=\frac{ee_{m}}{c}\widehat{R}$\ \cite{goldhaber,
Moura-Melo, Berard} and $\vec{R}$\ is the relative vector position between
the monopole and the electric charge. The vector $\vec{s}$\ is taken as an
angular momentum with independent degrees of freedom and must obey the
following classical Poisson-bracket relation%
\begin{equation}
\left\{ s_{i}\text{, }s_{j}\right\} =-\varepsilon _{ijk}s_{k}\text{.}
\end{equation}%
Observe that $H_{total}(\vec{p}$, $\vec{r})$\ is not spherically symmetric
due to the occurrence of $H_{string}$\ and even in the restricted case of $H(%
\vec{p}$, $\vec{r})$, the term $\vec{\nabla}\times \vec{A}_{\gamma }$\
breaks rotational invariance since 
\begin{equation}
\frac{d\vec{r}}{dt}\times \left( \vec{\nabla}\times \vec{A}_{\gamma }\right)
=\frac{d\vec{r}}{dt}\left( \vec{\nabla}\cdot \vec{A}_{\gamma }\right)
-\left( \frac{d\vec{r}}{dt}\cdot \vec{\nabla}\right) \vec{A}_{\gamma
}=-\left( \frac{d\vec{r}}{dt}\cdot \vec{\nabla}\right) \vec{A}_{\gamma }\neq
0\text{ in general.}
\end{equation}%
We made use of the transversality condition $\vec{\nabla}\cdot \vec{A}%
_{\gamma }=0$\ in computing $\frac{d\vec{r}}{dt}\times \left( \vec{\nabla}%
\times \vec{A}_{\gamma }\right) $\ \cite{ignatiev}. Furthermore, we
emphasize that we may obtain a spherically symmetric Hamiltonian provided
the auxiliary condition $\left( \frac{d\vec{r}}{dt}\right) _{k}\partial
_{k}\left( A_{\gamma }\right) _{j}$\ $=0$\ $\forall j=1$, $2$, $3$\ is
satisfied. Such condition is however unnecessary for our present analysis.

The Poisson brackets between two generic functions $f(\vec{p}$, $\vec{r}$, $%
t)$ and $g(\vec{p}$, $\vec{r}$, $t)$ of the dynamical variables $\vec{p}$
and $\vec{r}$, are defined as,%
\begin{equation}
\left\{ f(\vec{p}\text{, }\vec{r}\text{, }t)\text{, }g(\vec{p}\text{, }\vec{r%
}\text{, }t)\right\} \overset{\text{def}}{=}\underset{i}{\text{ }\sum }%
(\partial _{p_{i}}f\partial _{r_{i}}g-\partial _{r_{i}}f\partial _{p_{i}}g)
\end{equation}%
and the basic canonical Poisson bracket structure for the conjugate
variables is given by,%
\begin{equation}
\left\{ r_{i}\text{, }r_{j}\right\} =0\text{, }\left\{ r_{i}\text{, }%
p_{j}\right\} =-\delta _{ij}\text{, }\left\{ p_{i}\text{, }p_{j}\right\} =0%
\text{.}
\end{equation}%
Let us show explicitly that $\vec{J}$ is the generator of spatial rotations
so that we can safely define the rank of a tensor by studying its
transformation rules under such rotations. Let us prove,%
\begin{equation}
\left\{ J_{i}\text{, }J_{j}\right\} =-\varepsilon _{ijk}J_{k}\text{.}
\end{equation}%
Using the tensorial notation for the cross product appearing in the
definition of $\vec{J}$, and using the standard properties of a well-define
Poisson bracket structure, the brackets in equation $(22)$ become,%
\begin{eqnarray}
\left\{ J_{i}\text{, }J_{l}\right\} &=&\left\{ \varepsilon _{ijk}r_{j}p_{k}%
\text{, }\varepsilon _{lmn}r_{m}p_{n}\right\} -\left\{ \varepsilon
_{ijk}r_{j}p_{k}\text{, }\varepsilon _{lmn}r_{m}A_{n}\right\} +  \notag \\
&&-\left\{ \varepsilon _{ijk}r_{j}A_{k}\text{, }\varepsilon
_{lmn}r_{m}p_{n}\right\} +\left\{ \varepsilon _{ijk}r_{j}A_{k}\text{, }%
\varepsilon _{lmn}r_{m}A_{n}\right\} +\left\{ s_{i}\text{, }s_{l}\right\} 
\text{.}
\end{eqnarray}%
Using the basic canonical Poisson bracket structure expressed in $(21)$ and
the standard properties of Poisson brackets together with the following
identity,%
\begin{equation}
\varepsilon _{ijk}\varepsilon _{mlk}=\delta _{im}\delta _{jl}-\delta
_{il}\delta _{jm}
\end{equation}%
the first bracket on the rhs of $(23)$ becomes,%
\begin{equation}
\left\{ \varepsilon _{ijk}r_{j}p_{k}\text{, }\varepsilon
_{lmn}r_{m}p_{n}\right\} =r_{l}p_{i}-r_{i}p_{l}\text{.}\ 
\end{equation}%
Similarly, the second, the third and the fourth brackets on the rhs of $(23)$
become,%
\begin{equation}
-\left\{ \varepsilon _{ijk}r_{j}p_{k}\text{, }\varepsilon
_{lmn}r_{m}A_{n}\right\} =\delta _{il}r_{n}A_{n}-r_{l}A_{i}+\varepsilon
_{ijk}\varepsilon _{lmn}r_{m}p_{k}\left\{ A_{n}\text{, }r_{j}\right\}
\end{equation}%
\begin{equation}
-\left\{ \varepsilon _{ijk}r_{j}A_{k}\text{, }\varepsilon
_{lmn}r_{m}p_{n}\right\} =-\delta _{il}r_{k}A_{k}+r_{i}A_{l}+\varepsilon
_{ijk}\varepsilon _{lmn}r_{j}p_{n}\left\{ r_{m}\text{, }A_{k}\right\} \ 
\end{equation}%
\begin{equation}
\left\{ \varepsilon _{ijk}r_{j}A_{k}\text{, }\varepsilon
_{lmn}r_{m}A_{n}\right\} =-\varepsilon _{ijk}\varepsilon
_{lmn}r_{j}A_{n}\left\{ r_{m}\text{, }A_{k}\right\} -\varepsilon
_{ijk}\varepsilon _{lmn}r_{m}A_{k}\left\{ A_{n}\text{, }r_{j}\right\} \text{.%
}
\end{equation}%
The last bracket on the rhs of $(23)$ is given by $(18)$. Finally,
substituting these five brackets in the rhs of $(23)$ and ordering them
properly, the Poisson brackets of $\vec{J}$\ become,%
\begin{eqnarray}
\left\{ J_{i}\text{, }J_{l}\right\}
&=&(r_{l}p_{i}-r_{i}p_{l}-r_{l}A_{i}+r_{i}A_{l}-\varepsilon _{ilm}s_{m})\ + 
\notag \\
&&+\varepsilon _{ijk}\varepsilon _{lmn}\left[ r_{m}p_{k}\left\{ A_{n}\text{, 
}r_{j}\right\} -r_{j}p_{n}\left\{ r_{m}\text{, }A_{k}\right\} \right] + 
\notag \\
&&+\varepsilon _{ijk}\varepsilon _{lmn}\left[ r_{j}A_{n}\left\{ A_{k}\text{, 
}r_{m}\right\} -r_{m}A_{k}\left\{ A_{n}\text{, }r_{j}\right\} \right] \text{.%
}
\end{eqnarray}%
Because of the full antisymmetry of the Levi-Civita tensor,%
\begin{eqnarray}
\varepsilon _{ijk}\varepsilon _{lmn}r_{m}p_{k}\left\{ A_{n}\text{, }%
r_{j}\right\} -\varepsilon _{ijk}\varepsilon _{lmn}r_{j}p_{n}\left\{ A_{n}%
\text{, }r_{m}\right\} &=&  \notag \\
\left( \varepsilon _{ijk}\varepsilon _{lmn}-\varepsilon _{imn}\varepsilon
_{ljk}\right) r_{m}p_{k}\left\{ A_{n}\text{, }r_{j}\right\} &=&0\text{.}
\end{eqnarray}%
Therefore, equation $(29)$ becomes,%
\begin{eqnarray}
\left\{ J_{i}\text{, }J_{l}\right\}
&=&r_{l}p_{i}-r_{i}p_{l}-r_{l}A_{i}+r_{i}A_{l}-\varepsilon _{ilm}s_{m}= 
\notag \\
&=&-\varepsilon _{ilm}\left[ \varepsilon _{mnk}r_{n}(p_{k}-A_{k})+s_{m}%
\right] \text{.}
\end{eqnarray}%
Using equation $(24)$, we obtain%
\begin{equation}
-\varepsilon _{ilm}\varepsilon _{mnk}r_{n}p_{k}=(r_{l}p_{i}-r_{i}p_{l})\text{
and }\ \varepsilon _{ilm}\varepsilon
_{mnk}r_{n}A_{k}=-(r_{l}A_{i}-r_{i}A_{l})
\end{equation}%
and finally, 
\begin{equation}
\left\{ J_{i}\text{, }J_{l}\right\} =-\varepsilon _{ilm}J_{m}\text{.}
\end{equation}

At this point, we have all the elements to show the classical inconsistency
of the problem. Recall the kinetic momentum vector is defined as,%
\begin{equation}
\vec{P}\overset{\text{def}}{=}\vec{p}-\frac{e}{c}\vec{A}\text{, }\qquad \vec{%
A}=\vec{A}_{\gamma }+\vec{A}^{(Dirac)}\text{.}
\end{equation}%
Let us assume that there exist a well-defined Poisson bracket structure in
the classical theoretical setting in consideration. In particular, let us
assume a well-defined classical Poisson bracket structure among the vector
fields $\vec{J}$, $\vec{P}$, and $\vec{r}$, that is,%
\begin{equation}
\left\{ J_{i}\text{, }J_{j}\right\} =-\varepsilon _{ijk}J_{k}\text{, }%
\left\{ J_{i}\text{, }r_{j}\right\} =-\varepsilon _{ijk}r_{k}\text{, }%
\left\{ J_{i}\text{, }P_{j}\right\} =-\varepsilon _{ijk}P_{k}\text{.}
\end{equation}%
Being $\vec{J}$ the generator of rotations, it is required that any
arbitrary vector $\vec{v}$ must satisfy the following classical commutation
rules,%
\begin{equation}
\left\{ J_{i}\text{, }v_{j}\right\} =-\varepsilon _{ijk}v_{k}\text{.}
\end{equation}%
Therefore, let us study the transformation properties of the magnetic field
under spatial rotations. It must be,%
\begin{equation}
\left\{ J_{i}\text{, }B_{j}\right\} =-\varepsilon _{ijk}B_{k}\text{.}
\end{equation}%
In terms of the magnetic field decomposition, equation $(37)$ is equivalent
to,%
\begin{equation}
\left\{ J_{i}\text{, }B_{j}^{(Dirac)}\right\} =-\varepsilon
_{ijk}B_{k}^{(Dirac)}\text{ and }\left\{ J_{i}\text{, }(B_{\gamma
})_{j}\right\} =-\varepsilon _{ijk}(B_{\gamma })_{k}\text{.}
\end{equation}%
It is quite straightforward to check the validity of the first equation in $%
(38)$, as a matter of fact,%
\begin{eqnarray}
\left\{ J_{i}\text{, }B_{j}^{(Dirac)}\right\}  &=&\left\{ J_{i}\text{, }%
\frac{e_{m}}{r^{3}}r_{j}\right\} =\frac{e_{m}}{r^{3}}\left\{ J_{i}\text{, }%
r_{j}\right\} +\left\{ J_{i}\text{, }\frac{e_{m}}{r^{3}}\right\} r_{j} 
\notag \\
&=&-\varepsilon _{ijk}\frac{e_{m}}{r^{3}}r_{k}\equiv -\varepsilon
_{ijk}B_{k}^{(Dirac)}\text{.}
\end{eqnarray}%
Let us consider the validity of equation $(37)$, where the total magnetic
field $\vec{B}$\ is given by%
\begin{equation}
B_{j}\left( r\text{, }\theta \text{, }\varphi \right) =\varepsilon
_{jlm}\partial _{l}A_{m}\left( r\text{, }\theta \text{, }\varphi \right)
+e_{m}f_{j}\left( \vec{r}\right) \text{.}
\end{equation}%
By virtue of the "Dirac-veto", the magnetic field $B_{j}\left( r\text{, }%
\theta \text{, }\varphi \right) $\ "felt" by the electric charge reduces to%
\begin{equation}
B_{j}\left( r\text{, }\theta \text{, }\varphi \right) =\varepsilon
_{jlm}\partial _{l}A_{m}\left( r\text{, }\theta \text{, }\varphi \right) 
\text{.}
\end{equation}%
Fixing the constants $c$ and $e$ equal to one for the sake of convenience,
let us consider first the Poisson brackets of the kinetic momentum vector
components. Using $(21)$, the standard properties of Poisson brackets
together with equations $(24)$ and $(41)$, we obtain,%
\begin{equation}
\left\{ P_{i}\text{, }P_{j}\right\} =-\varepsilon _{ijk}B_{k}\text{.}
\end{equation}%
Multiplying both sides of $(42)$ by $\varepsilon _{ijn}$, we obtain%
\begin{equation}
\varepsilon _{ijn}\left\{ P_{i}\text{, }P_{j}\right\} =-\varepsilon
_{ijn}\varepsilon _{ijk}B_{k}=-2\delta _{nk}B_{k}=-2B_{n}
\end{equation}%
and therefore,%
\begin{equation}
B_{k}=-\frac{1}{2}\varepsilon _{ijk}\left\{ P_{i}\text{, }P_{j}\right\} 
\text{.}
\end{equation}%
Therefore, substituting $B_{k}$ of equation $(44)$ into $(37)$, we obtain%
\begin{equation}
\left\{ J_{i}\text{, }B_{j}\right\} =-\frac{1}{2}\varepsilon _{lmj}\left\{
J_{i}\text{, }\left\{ P_{l}\text{, }P_{m}\right\} \right\} \text{.}
\end{equation}%
The double commutator in equation $(45)$ cannot be calculated in a direct
way. However, because we are assuming the existence of a well-defined
Poisson bracket structure among the vectors $\vec{J}$, $\vec{B}$ and $\vec{r}
$, this double commutator can be evaluated by using the following Jacobi
identity,%
\begin{equation}
\left\{ J_{i}\text{, }\left\{ P_{l}\text{, }P_{m}\right\} \right\} +\left\{
P_{m}\text{, }\left\{ J_{i}\text{, }P_{l}\right\} \right\} +\left\{ P_{l}%
\text{, }\left\{ P_{m}\text{, }J_{i}\right\} \right\} =0\text{.}
\end{equation}%
Thus, using the fact that $\vec{J}$\ is the generator of rotations, that $%
\vec{P}$ transforms as a vector quantity under rotations, and using equation 
$(24)$, we obtain%
\begin{equation}
\left\{ J_{i}\text{, }\left\{ P_{l}\text{, }P_{m}\right\} \right\} =-\delta
_{il}B_{m}+\delta _{im}B_{l}\text{.}
\end{equation}%
Substituting equations $(44)$ into $(47)$, we obtain%
\begin{equation}
\left\{ J_{i}\text{, }B_{j}\right\} =-\varepsilon _{ijm}B_{m}\text{.}
\end{equation}%
Therefore, we have shown that in a pure classical theoretical framework
given by the Poisson brackets formalism, the commutation rule between the
generator of spatial rotations and the total magnetic field is expressed in $%
(48)$. Our last step is to calculate the Poisson brackets between $\vec{J}$
and the magnetic field $\vec{B}_{\gamma }$. Using equation $(6)$, standard
Poisson brackets properties and the fact that $\vec{J}$ is the generator of
rotations, these brackets become,%
\begin{equation}
\left\{ J_{i}\text{, }(\vec{B}_{\gamma })_{j}\right\}
_{Poisson}=-\varepsilon _{ijk}(B_{\gamma })_{k}+\left\{ J_{i}\text{, }%
b_{\gamma }^{(1)}\right\} r_{j}+\left\{ J_{i}\text{, }b_{\gamma
}^{(2)}\right\} n_{j}\text{.}
\end{equation}%
In order to have proper Poisson brackets, for each vectors $\widehat{n}$ and 
$\vec{r}$, the following relation must hold%
\begin{equation}
\left\{ J_{i}\text{, }b_{\gamma }^{(1)}\right\} r_{j}+\left\{ J_{i}\text{, }%
b_{\gamma }^{(2)}\right\} n_{j}=0\text{.}
\end{equation}%
Observe that the second Poisson bracket in the rhs of $(49)$ contains a term
quadratic in $n_{k}$,%
\begin{eqnarray}
\left\{ J_{i}\text{, }b_{\gamma }^{(2)}\right\} n_{j} &=&(\partial
_{p_{k}}J_{i})(\partial _{r_{k}}b_{\gamma }^{(2)})n_{j}=(\partial
_{p_{k}}J_{i})\left[ \partial _{r}b_{\gamma }^{(2)}\frac{r_{k}}{r}+\partial
_{\left( \vec{r}\cdot \widehat{n}\right) }b_{\gamma }^{(2)}n_{k}\right] n_{j}
\notag \\
&=&\frac{1}{r}\partial _{p_{k}}J_{i}\ \partial _{r}b_{\gamma
}^{(2)}r_{k}n_{j}+\partial _{p_{k}}J_{i}\partial _{\left( \vec{r}\cdot 
\widehat{n}\right) }b_{\gamma }^{(2)}n_{k}n_{j}\text{.}
\end{eqnarray}%
Since, the proper Poisson brackets should be linear in $n_{k}$, we require%
\begin{equation}
\partial _{\left( \vec{r}\cdot \widehat{n}\right) }b_{\gamma }^{(2)}=0\text{.%
}
\end{equation}%
There is no way to cancel out this term in $(49)$, then it must be,%
\begin{equation}
b_{\gamma }^{(2)}=0\text{.}
\end{equation}%
We now consider the first Poisson bracket on the rhs of $(49)$. Because of
the anti-symmetry in the indices $i$ and $j$ of the term $\varepsilon
_{ijk}(B_{\gamma })_{k}$, it must be%
\begin{equation}
\left\{ J_{i}\text{, }b_{\gamma }^{(1)}\right\} r_{j}+\left\{ J_{j}\text{, }%
b_{\gamma }^{(1)}\right\} r_{i}=0
\end{equation}%
that is,%
\begin{equation}
\left\{ J_{i}\text{, }b_{\gamma }^{(1)}\right\} r_{i}=0\text{.}
\end{equation}%
Explicitly, equation $(55)$ becomes,%
\begin{eqnarray}
0 &=&(\partial _{p_{k}}J_{i})(\partial _{r_{k}}b_{\gamma
}^{(1)})r_{i}=(\partial _{p_{k}}J_{i})\left[ \partial _{r}b_{\gamma }^{(1)}%
\frac{r_{k}}{r}+\partial _{\left( \vec{r}\cdot \overset{\wedge }{n}\right)
}b_{\gamma }^{(1)}n_{k}\right] r_{i}=  \notag \\
&=&\frac{1}{r}(\partial _{p_{k}}J_{i})(\ \partial _{r}b_{\gamma
}^{(1)})r_{k}r_{i}+(\partial _{p_{k}}J_{i})(\partial _{\left( \vec{r}\cdot 
\overset{\wedge }{n}\right) }b_{\gamma }^{(1)})n_{k}r_{i}\text{.}
\end{eqnarray}%
We neglect the quadratic term in $r_{k}$ in equation $(56)$ since this term
has no analog in the proper Poisson brackets. Then, we have%
\begin{equation}
\partial _{\left( \vec{r}\cdot \widehat{n}\right) }b_{\gamma }^{(1)}=0\text{.%
}
\end{equation}%
Recalling that%
\begin{equation}
\widehat{n}=-\widehat{z}=-\left\{ \cos (\theta )\widehat{r}-\sin (\theta )%
\widehat{\theta }\right\} =-\cos (\theta )\widehat{r}+\sin (\theta )\widehat{%
\theta }
\end{equation}%
then,%
\begin{equation}
\widehat{n}\cdot \widehat{r}=-\cos (\theta )=\theta -\text{dependent.}
\end{equation}%
Therefore, equation $(57)$ is satisfied by an arbitrary scalar function $%
b_{\gamma }(r)$. As a consequence, the magnetic field $\vec{B}_{\gamma }$ is
not $\theta -$dependent (in a more general situation in which $\widehat{n}$
is not along the z-axis, we would conclude that the magnetic field is not $%
(\theta ,\varphi )-$dependent). $\vec{B}_{\gamma }$ must be a spherically
symmetric field whose general expression is the following,%
\begin{equation}
\vec{B}_{\gamma }\left( \vec{r}\right) =B_{\gamma }(r)\widehat{r}\text{.}
\end{equation}%
In conclusion, in order to have a well-defined classical Poisson bracket
structure in the problem under investigation, one must deal with diffuse
magnetic field solutions exhibiting spherical symmetry. However, those very
same solutions are not compatible with massive classical electrodynamics
with magnetic monopoles. This result means that it is not possible to
formulate a consistent non-relativistic classical theory describing the
finite-range electromagnetic interaction between a point-like electric
charge and a fixed Dirac monopole without a "visible" string. In other
words, there is no way to construct a consistent Lie algebra in our
classical framework and this leads to the conclusion that there is no
angular momentum to be quantized in order to give the Dirac quantization
rule. This fact points out that the string attached to the monopole is
visible and there is no way to make it invisible when considering
finite-range electromagnetic interactions in a pure classical framework. The
Dirac string must assume dynamical significance if the photon has a
non-vanishing mass and its dynamical evolution may play a significant role
in a quantum description of the Dirac theory. In conclusion, we have shown
that it is not possible to construct a non-relativistic classical theory of
"true" Dirac monopoles (invisible string, "monopole without a string") and
massive photons unless the string attached to the monopole is treated as an
independent dynamical quantity. An important feature of our approach is that
we do not use any kind of semiclassical approximation or limiting procedure
for $\hbar \rightarrow 0$.

\appendix

\section{The Generator of Spatial Rotations}

We show that $\vec{J}$ is the generator of spatial rotations, that is,%
\begin{equation}
\left\{ J_{i}\text{, }J_{j}\right\} =-\varepsilon _{ijk}J_{k}\text{.}
\end{equation}%
Notice that,%
\begin{eqnarray}
\left\{ J_{i}\text{, }J_{l}\right\} &=&\left\{ \varepsilon _{ijk}r_{j}\left(
p_{k}-A_{k}\right) +s_{i}\text{, }\varepsilon _{lmn}r_{m}\left(
p_{n}-A_{n}\right) +s_{l}\right\}  \notag \\
&=&\left\{ \varepsilon _{ijk}r_{j}p_{k}-\varepsilon _{ijk}r_{j}A_{k}+s_{i}%
\text{, }\varepsilon _{lmn}r_{m}p_{n}-\varepsilon
_{lmn}r_{m}A_{n}+s_{l}\right\}  \notag \\
&=&\left\{ \varepsilon _{ijk}r_{j}p_{k}\text{, }\varepsilon
_{lmn}r_{m}p_{n}\right\} -\left\{ \varepsilon _{ijk}r_{j}p_{k}\text{, }%
\varepsilon _{lmn}r_{m}A_{n}\right\} -\left\{ \varepsilon _{ijk}r_{j}A_{k}%
\text{, }\varepsilon _{lmn}r_{m}p_{n}\right\} +  \notag \\
&&+\left\{ \varepsilon _{ijk}r_{j}A_{k}\text{, }\varepsilon
_{lmn}r_{m}A_{n}\right\} +\left\{ s_{i}\text{, }s_{l}\right\} \text{.}
\end{eqnarray}%
Therefore there are five Poisson brackets to be calculated. Consider the
first one,%
\begin{eqnarray}
\left\{ \varepsilon _{ijk}r_{j}p_{k}\text{, }\varepsilon
_{lmn}r_{m}p_{n}\right\} &=&\varepsilon _{ijk}\varepsilon _{lmn}\left\{
r_{j}p_{k}\text{, }r_{m}p_{n}\right\} =\varepsilon _{ijk}\varepsilon _{lmn}%
\left[ r_{j}\left\{ p_{k}\text{, }r_{m}p_{n}\right\} +\left\{ r_{j}\text{, }%
r_{m}p_{n}\right\} p_{k}\right]  \notag \\
&=&\varepsilon _{ijk}\varepsilon _{lmn}\left[ -r_{j}\left\{ r_{m}p_{n}\text{%
, }p_{k}\right\} -\left\{ r_{m}p_{n}\text{, }r_{j}\right\} p_{k}\right] 
\notag \\
&=&\varepsilon _{ijk}\varepsilon _{lmn}\left[ -r_{j}\left( r_{m}\left\{ p_{n}%
\text{, }p_{k}\right\} +\left\{ r_{m}\text{, }p_{k}\right\} p_{n}\right) %
\right] +\varepsilon _{ijk}\varepsilon _{lmn}\left[ -\left( r_{m}\left\{
p_{n}\text{, }r_{j}\right\} +\left\{ r_{m}\text{, }r_{j}\right\}
p_{n}\right) p_{k}\right]  \notag \\
&=&\varepsilon _{ijk}\varepsilon _{lmn}\left[ \delta _{mk}r_{j}p_{n}-\delta
_{nj}r_{m}p_{k}\right] =\varepsilon _{ijk}\varepsilon _{lmn}\delta
_{mk}r_{j}p_{n}-\varepsilon _{ijk}\varepsilon _{lmn}\delta _{nj}r_{m}p_{k} 
\notag \\
&=&\varepsilon _{ijk}\varepsilon _{lkn}r_{j}p_{n}-\varepsilon
_{ink}\varepsilon _{lmn}r_{m}p_{k}=-\varepsilon _{ijk}\varepsilon
_{lnk}r_{j}p_{n}+\varepsilon _{ikn}\varepsilon _{lmn}r_{m}p_{k}  \notag \\
&=&-\left( \delta _{il}\delta _{jn}-\delta _{in}\delta _{jl}\right)
r_{j}p_{n}+\left( \delta _{il}\delta _{km}-\delta _{im}\delta _{lk}\right)
r_{m}p_{k}  \notag \\
&=&-\delta _{il}\delta _{jn}r_{j}p_{n}+\delta _{in}\delta
_{jl}r_{j}p_{n}+\delta _{il}\delta _{km}r_{m}p_{k}-\delta _{im}\delta
_{lk}r_{m}p_{k}  \notag \\
&=&-\delta _{il}r_{n}p_{n}+r_{l}p_{i}+\delta
_{il}r_{k}p_{k}-r_{i}p_{l}=r_{l}p_{i}-r_{i}p_{l}
\end{eqnarray}%
thus,%
\begin{equation}
\left\{ \varepsilon _{ijk}r_{j}p_{k}\text{, }\varepsilon
_{lmn}r_{m}p_{n}\right\} =r_{l}p_{i}-r_{i}p_{l}\text{.}
\end{equation}%
Consider the second bracket,%
\begin{eqnarray}
-\left\{ \varepsilon _{ijk}r_{j}p_{k}\text{, }\varepsilon
_{lmn}r_{m}A_{n}\right\} &=&-\varepsilon _{ijk}\varepsilon _{lmn}\left\{
r_{j}p_{k}\text{, }r_{m}A_{n}\right\}  \notag \\
&=&-\varepsilon _{ijk}\varepsilon _{lmn}\left[ r_{j}\left\{ p_{k}\text{, }%
r_{m}A_{n}\right\} +\left\{ r_{j}\text{, }r_{m}A_{n}\right\} p_{k}\right] 
\notag \\
&=&-\varepsilon _{ijk}\varepsilon _{lmn}\left[ -r_{j}\left\{ r_{m}A_{n}\text{%
, }p_{k}\right\} -\left\{ r_{m}A_{n}\text{, }r_{j}\right\} p_{k}\right] 
\notag \\
&=&-\varepsilon _{ijk}\varepsilon _{lmn}\left[ -r_{j}r_{m}\left\{ A_{n}\text{%
, }p_{k}\right\} -r_{j}\left\{ r_{m}\text{, }p_{k}\right\} A_{n}\right] + 
\notag \\
&&-\varepsilon _{ijk}\varepsilon _{lmn}\left[ -r_{m}\left\{ A_{n}\text{, }%
r_{j}\right\} p_{k}-\left\{ r_{m}\text{, }r_{j}\right\} A_{n}p_{k}\right] 
\notag \\
&=&-\varepsilon _{ijk}\varepsilon _{lmn}\left[ \delta
_{mk}r_{j}A_{n}-r_{m}p_{k}\left\{ A_{n}\text{, }r_{j}\right\} \right]  \notag
\\
&=&-\varepsilon _{ijk}\varepsilon _{lkn}r_{j}A_{n}+\varepsilon
_{ijk}\varepsilon _{lmn}r_{m}p_{k}\left\{ A_{n}\text{, }r_{j}\right\}  \notag
\\
&=&\varepsilon _{ijk}\varepsilon _{l,n,k}r_{j}A_{n}+\varepsilon
_{ijk}\varepsilon _{lmn}r_{m}p_{k}\left\{ A_{n}\text{, }r_{j}\right\}  \notag
\\
&=&\left( \delta _{il}\delta _{jn}-\delta _{in}\delta _{jl}\right)
r_{j}A_{n}+\varepsilon _{ijk}\varepsilon _{lmn}r_{m}p_{k}\left\{ A_{n}\text{%
, }r_{j}\right\}  \notag \\
&=&\delta _{il}\delta _{jn}r_{j}A_{n}-\delta _{in}\delta
_{jl}r_{j}A_{n}+\varepsilon _{ijk}\varepsilon _{lmn}r_{m}p_{k}\left\{ A_{n}%
\text{, }r_{j}\right\}  \notag \\
&=&\delta _{il}r_{n}A_{n}-r_{l}A_{i}+\varepsilon _{ijk}\varepsilon
_{lmn}r_{m}p_{k}\left\{ A_{n}\text{, }r_{j}\right\}
\end{eqnarray}%
thus,%
\begin{equation}
-\left\{ \varepsilon _{ijk}r_{j}p_{k}\text{, }\varepsilon
_{lmn}r_{m}A_{n}\right\} =\delta _{il}r_{n}A_{n}-r_{l}A_{i}+\varepsilon
_{ijk}\varepsilon _{lmn}r_{m}p_{k}\left\{ A_{n}\text{, }r_{j}\right\} \text{.%
}
\end{equation}%
Using the standard canonical algebra, the third bracket becomes,%
\begin{equation}
-\left\{ \varepsilon _{ijk}r_{j}A_{k}\text{, }\varepsilon
_{lmn}r_{m}p_{n}\right\} =-\delta _{il}r_{k}A_{k}+r_{i}A_{l}+\varepsilon
_{ijk}\varepsilon _{lmn}r_{j}p_{n}\left\{ r_{m}\text{, }A_{k}\right\} \text{.%
}
\end{equation}%
For the fourth bracket, we obtain%
\begin{eqnarray}
\left\{ \varepsilon _{ijk}r_{j}A_{k}\text{, }\varepsilon
_{lmn}r_{m}A_{n}\right\} &=&\varepsilon _{ijk}\varepsilon _{lmn}\left\{
r_{j}A_{k}\text{, }r_{m}A_{n}\right\}  \notag \\
&=&\varepsilon _{ijk}\varepsilon _{lmn}\left[ r_{j}\left\{ A_{k}\text{, }%
r_{m}A_{n}\right\} +\left\{ r_{j}\text{, }r_{m}A_{n}\right\} A_{k}\right] 
\notag \\
&=&\varepsilon _{ijk}\varepsilon _{lmn}\left[ -r_{j}\left\{ r_{m}A_{n}\text{%
, }A_{k}\right\} -\left\{ r_{m}A_{n}\text{, }r_{j}\right\} A_{k}\right] 
\notag \\
&=&\varepsilon _{ijk}\varepsilon _{lmn}\left[ -r_{j}\left\{ r_{m}\text{, }%
A_{k}\right\} A_{n}-r_{m}\left\{ A_{n}\text{, }r_{j}\right\} A_{k}\right] 
\notag \\
&=&-\varepsilon _{ijk}\varepsilon _{lmn}r_{j}A_{n}\left\{ r_{m}\text{, }%
A_{k}\right\} -\varepsilon _{ijk}\varepsilon _{lmn}r_{m}A_{k}\left\{ A_{n}%
\text{, }r_{j}\right\} \text{.}
\end{eqnarray}%
For the last bracket, let us remind that the vector $s$ is such the Poisson
brackets of its components satisfy equation $(18)$. In conclusion, using
equations $(A4)$, $(A6)$, $(A7)$, $(A8)$ and using the commutation rules of
the classical spin, equation $(A2)$ becomes,%
\begin{eqnarray}
\left\{ J_{i}\text{, }J_{l}\right\} &=&r_{l}p_{i}-r_{i}p_{l}+\delta
_{il}r_{n}A_{n}-r_{l}A_{i}+\varepsilon _{ijk}\varepsilon
_{lmn}r_{m}p_{k}\left\{ A_{n}\text{, }r_{j}\right\} -\delta _{il}r_{k}A_{k}+
\notag \\
&&+r_{i}A_{l}+\varepsilon _{ijk}\varepsilon _{lmn}r_{j}p_{n}\left\{ r_{m}%
\text{, }A_{k}\right\} -\varepsilon _{ijk}\varepsilon
_{lmn}r_{j}A_{n}\left\{ r_{m}\text{, }A_{k}\right\} +  \notag \\
&&-\varepsilon _{ijk}\varepsilon _{lmn}r_{m}A_{k}\left\{ A_{n}\text{, }%
r_{j}\right\} -\varepsilon _{ilm}s_{m}  \notag \\
&=&(r_{l}p_{i}-r_{i}p_{l}-r_{l}A_{i}+r_{i}A_{l}-\varepsilon _{ilm}s_{m})+ 
\notag \\
&&+(\varepsilon _{ijk}\varepsilon _{lmn}\left[ r_{m}p_{k}\left\{ A_{n}\text{%
, }r_{j}\right\} -r_{j}p_{n}\left\{ r_{m}\text{, }A_{k}\right\}
+r_{j}A_{n}\left\{ A_{k}\text{, }r_{m}\right\} -r_{m}A_{k}\left\{ A_{n}\text{%
, }r_{j}\right\} \right] )\text{.}
\end{eqnarray}%
Notice that,%
\begin{equation}
\varepsilon _{ijk}\varepsilon _{lmn}r_{m}p_{k}\left\{ A_{n}\text{, }%
r_{j}\right\} -\varepsilon _{ijk}\varepsilon _{lmn}r_{j}p_{n}\left\{ A_{n}%
\text{, }r_{m}\right\} =\left( \varepsilon _{ijk}\varepsilon
_{lmn}-\varepsilon _{imn}\varepsilon _{ljk}\right) r_{m}p_{k}\left\{ A_{n}%
\text{, }r_{j}\right\} =0\text{.}
\end{equation}%
If $i=l$, then,%
\begin{equation}
\varepsilon _{ijk}\varepsilon _{lmn}-\varepsilon _{imn}\varepsilon
_{ljk}=\varepsilon _{ijk}\varepsilon _{imn}-\varepsilon _{imn}\varepsilon
_{ijk}\equiv 0\text{.}
\end{equation}%
If $i\neq l$, let us say $i=1$ and $l=2$, then%
\begin{equation}
\varepsilon _{ijk}\varepsilon _{lmn}-\varepsilon _{imn}\varepsilon
_{ljk}=\varepsilon _{1jk}\varepsilon _{2mn}-\varepsilon _{1mn}\varepsilon
_{2jk}\text{.}
\end{equation}%
Therefore, the possible non-vanishing pieces are:%
\begin{equation}
\varepsilon _{123}\varepsilon _{213}-\varepsilon _{213}\varepsilon
_{123}\equiv 0\text{, }\varepsilon _{132}\varepsilon _{231}-\varepsilon
_{231}\varepsilon _{132}\equiv 0\text{, }\varepsilon _{132}\varepsilon
_{213}-\varepsilon _{231}\varepsilon _{123}\equiv 0\text{, etc. etc.}
\end{equation}%
Therefore, equation $(A9)$ becomes,%
\begin{eqnarray}
\left\{ J_{i}\text{, }J_{l}\right\}
&=&(r_{l}p_{i}-r_{i}p_{l}-r_{l}A_{i}+r_{i}A_{l}-\varepsilon _{ilm}s_{m}) 
\notag \\
&=&-\varepsilon _{ilm}\left[ \varepsilon _{mnk}r_{n}(p_{k}-A_{k})+s_{m}%
\right]  \notag \\
&=&-\varepsilon _{ilm}J_{m}\text{.}
\end{eqnarray}%
Indeed,%
\begin{eqnarray}
-\varepsilon _{ilm}\varepsilon _{mnk}r_{n}p_{k} &=&-\varepsilon
_{ilm}\varepsilon _{kmn}r_{n}p_{k}=-\varepsilon _{ilm}\varepsilon
_{nkm}r_{n}p_{k}=  \notag \\
&=&\left( \delta _{in}\delta _{lk}-\delta _{ik}\delta _{l,n}\right)
r_{n}p_{k}=-\delta _{in}\delta _{lk}r_{n}p_{k}+\delta _{ik}\delta
_{l,n}r_{n}p_{k}  \notag \\
&=&-r_{i}p_{l}+r_{l}p_{i}=(r_{l}p_{i}-r_{i}p_{l})
\end{eqnarray}%
and,%
\begin{equation}
\varepsilon _{ilm}\varepsilon
_{mnk}r_{n}A_{k}=r_{i}A_{l}+r_{l}A_{i}=-(r_{l}A_{i}-r_{i}A_{l})\text{.}
\end{equation}%
This concludes our proof.

\section{The Jacobi Identity}

Consider the kinetic momentum vector,%
\begin{equation}
\vec{P}\overset{\text{def}}{=}\vec{p}-\frac{e}{c}\vec{A}\text{, }\vec{A}=%
\vec{A}_{\gamma }+\vec{A}^{(Dirac)}\text{.}
\end{equation}%
Consider the Poisson bracket of the kinetic momentum vector components,%
\begin{eqnarray}
\left\{ P_{i}\text{, }P_{j}\right\}  &=&\left\{ p_{i}-A_{i}\text{, }%
p_{j}-A_{j}\right\} =  \notag \\
&=&\left\{ p_{i}\text{, }p_{j}\right\} -\left\{ p_{i}\text{, }A_{j}\right\}
-\left\{ A_{i}\text{, }p_{j}\right\} +\left\{ A_{i}\text{, }A_{j}\right\}
=\left\{ A_{j}\text{, }p_{i}\right\} -\left\{ A_{i}\text{, }p_{j}\right\}  
\notag \\
&=&\left\{ A_{j}\text{, }p_{i}\right\} -\left\{ A_{i}\text{, }p_{j}\right\}
=-\partial iA_{j}+\partial _{j}A_{i}=-(\partial iA_{j}-\partial _{j}A_{i}) 
\notag \\
&=&-\varepsilon _{ijk}B_{k}
\end{eqnarray}%
where%
\begin{equation}
B_{j}=\varepsilon _{jlm}\partial _{l}A_{m}\text{.}
\end{equation}%
Using the fact that $\left\{ J_{i},B_{j}\right\} =-\varepsilon _{ijk}B_{k}$
and the identity $\varepsilon _{ijk}\varepsilon _{mlk}=\delta _{il}\delta
_{jm}-\delta _{im}\delta _{jl}$, it follows that,%
\begin{eqnarray}
\varepsilon _{ijk}B_{k} &=&\varepsilon _{ijk}\varepsilon _{klm}\partial
_{l}A_{m}=\varepsilon _{ijk}\varepsilon _{mkl}\partial
_{l}A_{m}=-\varepsilon _{ijk}\varepsilon _{mlk}\partial _{l}A_{m}  \notag \\
&=&-(\delta _{im}\delta _{jl}-\delta _{il}\delta _{jm})\partial
_{l}A_{m}=-\delta _{im}\delta _{jl}\partial _{l}A_{m}+\delta _{il}\delta
_{jm}\partial _{l}A_{m}  \notag \\
&=&-\delta _{im}\partial _{j}A_{m}+\delta _{il}\partial _{l}A_{j}=\partial
_{i}A_{j}-\partial _{j}A_{i}\text{.}
\end{eqnarray}%
Using equation $(B2)$, we obtain%
\begin{equation}
\varepsilon _{ijn}\left\{ P_{i}\text{, }P_{j}\right\} =-\varepsilon
_{ijn}\varepsilon _{ijk}B_{k}=-2\delta _{nk}B_{k}=-2B_{n}\text{.}
\end{equation}%
Thus,%
\begin{equation}
B_{k}=-\frac{1}{2}\varepsilon _{ijk}\left\{ P_{i}\text{, }P_{j}\right\} 
\text{.}
\end{equation}%
Finally, let us focus on the following Poisson bracket,%
\begin{equation}
\left\{ J_{i}\text{, }B_{j}\right\} =\left\{ J_{i}\text{,}-\frac{1}{2}%
\varepsilon _{lmj}\left\{ P_{l}\text{, }P_{m}\right\} \right\} =-\frac{1}{2}%
\varepsilon _{lmj}\left\{ J_{i}\text{, }\left\{ P_{l}\text{, }P_{m}\right\}
\right\} \text{.}
\end{equation}%
Using the Jacobi identity,%
\begin{equation}
\left\{ J_{i}\text{, }\left\{ P_{l}\text{, }P_{m}\right\} \right\} +\left\{
P_{m}\text{, }\left\{ J_{i}\text{, }P_{l}\right\} \right\} +\left\{ P_{l}%
\text{, }\left\{ P_{m}\text{, }J_{i}\right\} \right\} =0
\end{equation}%
we obtain%
\begin{eqnarray}
\left\{ J_{i}\text{, }\left\{ P_{l}\text{, }P_{m}\right\} \right\} 
&=&-\left\{ P_{m}\text{, }\left\{ J_{i}\text{, }P_{l}\right\} \right\}
-\left\{ P_{l}\text{, }\left\{ P_{m}\text{, }J_{i}\right\} \right\} =\left\{
P_{l}\text{, }\left\{ J_{i}\text{, }P_{m}\right\} \right\} -\left\{ P_{m}%
\text{, }\left\{ J_{i}\text{, }P_{l}\right\} \right\}   \notag \\
&=&\left\{ P_{l}\text{, }-\varepsilon _{imk}P_{k}\right\} -\left\{ P_{m}%
\text{, }-\varepsilon _{ilk}P_{k}\right\} =-\varepsilon _{imk}\left\{ P_{l}%
\text{, }P_{k}\right\} +\varepsilon _{ilk}\left\{ P_{m}\text{, }%
P_{k}\right\}   \notag \\
&=&-\varepsilon _{imk}(-\varepsilon _{lkq}B_{q})+\varepsilon
_{ilk}(-\varepsilon _{mkq}B_{q})=\varepsilon _{imk}\varepsilon
_{lkq}B_{q}-\varepsilon _{ilk}\varepsilon _{mkq}B_{q}  \notag \\
&=&-\varepsilon _{imk}\varepsilon _{lqk}B_{q}+\varepsilon _{ilk}\varepsilon
_{mqk}B_{q}=-(\delta _{il}\delta _{mq}-\delta _{iq}\delta
_{ml})B_{q}+(\delta _{im}\delta _{lq}-\delta _{iq}\delta _{lm})B_{q}  \notag
\\
&=&-\delta _{il}\delta _{mq}B_{q}+\delta _{iq}\delta _{ml}B_{q}+\delta
_{im}\delta _{lq}B_{q}-\delta _{iq}\delta _{lm}B_{q}  \notag \\
&=&-\delta _{il}B_{m}+\delta _{ml}B_{i}+\delta _{im}B_{l}-\delta
_{lm}B_{i}=-\delta _{il}B_{m}+\delta _{im}B_{l}\text{.}
\end{eqnarray}%
Then, using equations $(B6)$ and $(B9)$, we obtain%
\begin{eqnarray}
\left\{ J_{i}\text{, }B_{j}\right\}  &=&-\frac{1}{2}\varepsilon
_{lmj}(-\delta _{il}B_{m}+\delta _{im}B_{l})=\frac{1}{2}\varepsilon
_{lmj}\delta _{il}B_{m}-\frac{1}{2}\varepsilon _{lmj}\delta _{im}B_{l} 
\notag \\
&=&\frac{1}{2}\varepsilon _{imj}B_{m}-\frac{1}{2}\varepsilon _{lij}B_{l}=-%
\frac{1}{2}\varepsilon _{ijm}B_{m}-\frac{1}{2}\varepsilon _{mij}B_{m}  \notag
\\
&=&-\frac{1}{2}\varepsilon _{ijm}B_{m}-\frac{1}{2}\varepsilon
_{ijm}B_{m}=-\varepsilon _{ijm}B_{m}\text{.}
\end{eqnarray}%
We have shown that in a pure classical theoretical framework given by the
Poisson brackets formalism, the commutation rule between the generator of
spatial rotations and the total magnetic field is,%
\begin{equation}
\left\{ J_{i},B_{j}\right\} =i\varepsilon _{ijk}B_{k}\text{.}
\end{equation}

\begin{acknowledgments}
We are grateful to A. Caticha and J. Kimball for useful comments.
\end{acknowledgments}

\end{document}